\documentclass[10pt, conference,]{IEEEtran}
\IEEEoverridecommandlockouts
\usepackage{cite}
\usepackage{amsmath,amssymb,amsfonts}
\usepackage{algorithmic}
\usepackage{graphicx}
\usepackage{textcomp}
\usepackage{xcolor}
\usepackage{subfigure}
\usepackage[ruled,linesnumbered, vlined]{algorithm2e}
\usepackage{enumitem}
\usepackage{multirow} 
\usepackage{ulem}
\usepackage{amssymb}
\usepackage{bbding}
\usepackage{pifont}
\usepackage{tikz}
\newcommand*\circled[1]{\tikz[baseline=(char.base)]{
            \node[shape=circle,draw,white, fill=black,inner sep=1pt] (char) {#1};}}
\newcommand*\squared[1]{\tikz[baseline=(char.base)]{
            \node[shape=rectangle,draw,white, fill=black, minimum size=10pt, inner sep=1pt] (char) {#1};}}

\def\BibTeX{{\rm B\kern-.05em{\sc i\kern-.025em b}\kern-.08em
    T\kern-.1667em\lower.7ex\hbox{E}\kern-.125emX}}

\begin{document}

% \title{Galaxy: Enhancing Edge Intelligence by Distributed In-situ Transformer Inference}
\title{Galaxy: A Resource-Efficient Collaborative Edge AI System for In-situ Transformer Inference}

\author{
    \IEEEauthorblockN{Shengyuan Ye$^\blacklozenge$, Jiangsu Du$^\blacklozenge$, Liekang Zeng$^\blacklozenge$$^\lozenge$, Wenzhong Ou$^\blacklozenge$, Xiaowen Chu$^\blacktriangle$$^\vartriangle$, Yutong Lu$^\blacklozenge$, Xu Chen$^\blacklozenge$
    }
    \IEEEauthorblockA{$^\blacklozenge$School of Computer Science and Engineering, Sun Yat-sen University, Guangzhou, China}
    \IEEEauthorblockA{$^\lozenge$IoT Thrust and Research Center for Digital World with Intelligent Things, HKUST (Guangzhou), Guangzhou, China}
    \IEEEauthorblockA{$^\blacktriangle$Data Science and Analytics Thrust, HKUST (Guangzhou), Guangzhou, China}
    \IEEEauthorblockA{$^\vartriangle$Department of Computer Science and Engineering, HKUST, Hong Kong SAR, China}
    \IEEEauthorblockA{\{yeshy8, zenglk3, ouwzh3\}@mail2.sysu.edu.cn, \{dujiangsu, luyutong, chenxu35\}@mail.sysu.edu.cn, xwchu@ust.hk}
    % \thanks{$^*$Corresponding authors: Xu Chen and Jiangsu Du}
}

\maketitle

\begin{abstract}
Transformer-based models have unlocked a plethora of powerful intelligent applications at the edge, such as voice assistant in smart home.
Traditional deployment approaches offload the inference workloads to the remote cloud server, which would induce substantial pressure on the backbone network as well as raise users' privacy concerns.
To address that, in-situ inference has been recently recognized for edge intelligence, but it still confronts significant challenges stemming from the conflict between intensive workloads and limited on-device computing resources.
In this paper, we leverage our observation that many edge environments usually comprise a rich set of accompanying trusted edge devices with idle resources and propose Galaxy, a collaborative edge AI system that breaks the resource walls across heterogeneous edge devices for efficient Transformer inference acceleration.
Galaxy introduces a novel hybrid model parallelism to orchestrate collaborative inference, along with a heterogeneity-aware parallelism planning for fully exploiting the resource potential.
Furthermore, Galaxy devises a tile-based fine-grained overlapping of communication and computation to mitigate the impact of tensor synchronizations on inference latency under bandwidth-constrained edge environments.
Extensive evaluation based on prototype implementation demonstrates that Galaxy remarkably outperforms state-of-the-art approaches under various edge environment setups, achieving up to $2.5\times$ end-to-end latency reduction.
\end{abstract}

\section{Introduction}
Transformer-based models \cite{devlin2018bert, radford2018improving} have achieved superior performance in the field of Natural Language Processing (NLP) and driven increasing intelligent applications at the network edge.
In edge intelligent applications, such as AI assistants in smart homes \cite{king2023sasha} and voice-controlled robots in smart factories \cite{vemprala2023chatgpt}, single-shot inference (referring to single-command requests) tasks are prevalent, necessitating efficient and low-latency inference for seamless user interactions.
Currently, most Transformer-based intelligent applications heavily depend on cloud services, with the actual inference of large-scale Transformer-based models taking place in the cloud \cite{li2023alpaserve, fang2021turbotransformers}. At the edge, only a proxy daemon is deployed to forward user requests \cite{king2023sasha}. However, the cloud-assisted approaches suffer from following issues:
(1) Quality-of-Service may suffer due to unreliable and delay-prone wide-area network (WAN) connections between edge devices and remote clouds \cite{zhou2019edge}.
(2) Inference requests from numerous edge clients can impose significant pressure on both the backbone network and datacenters.
(3) The sensory data in smart applications can contain highly sensitive or private information. Transferring these data to the remote cloud owned by commercial companies inevitably raises users’ privacy concerns \cite{voigt2017eu}.

To address that, in-situ inference \cite{jia2022codl,zeng2020coedge} on edge devices without remote assistance, which keeps data locally and avoids network transmission, has been recognized as a promising paradigm for intelligent applications at the edge. 
However, the computation-intensive and resource-hungry nature of Transformer inference presents significant challenges for resource-constrained edge devices \cite{ye2022eco}. As we will show in \S \ref{sec:motivation-edge}, inference on the Bert-L model \cite{radford2019language-gpt2} in an off-the-shelf edge device imposes a minimum available memory space of almost 700MB, while taking $121\times$ longer latency than that in a datacenter GPU. These results demonstrate the fundamental contradiction between intensive Transformer inference workload and constrained onboard resources. To tackle these challenges, existing arts explore to design sophisticated scheduling mechanisms to leverage the resource potential of edge devices \cite{wang2021asymo, wang2019high,jia2022codl,kim2019mulayer}, but are still bottlenecked by the limited onboard resource of a single device.

\begin{figure}[t!]
    \setlength{\abovecaptionskip}{-0.1cm}
    \centering
    \includegraphics[width=0.9\linewidth]{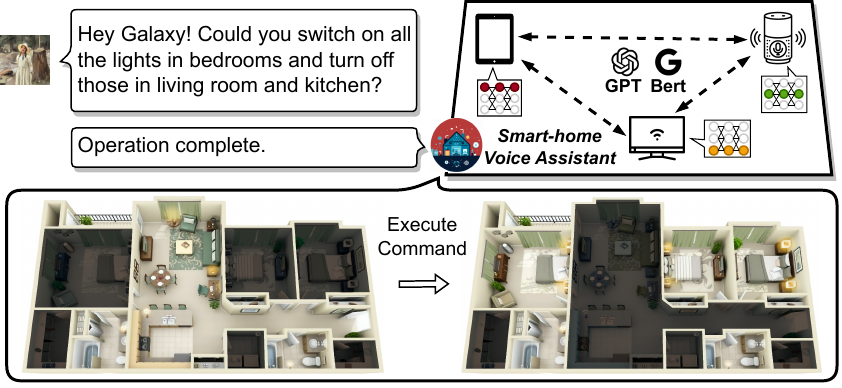}
    % \caption{Galaxy System Overview.}
    \caption{AI assistant in smart home scenario empowered by Galaxy.}
    \label{fig:intro}
    \vspace{-15pt}
\end{figure}

Alternatively, we observe that prevalent edge environments like smart homes usually comprise a rich set of trusted idle devices in physical proximity \cite{zeng2020coedge, zhao2018deepthings}. This motivates us to regard vicinal available edge devices as a resource augmentation and collaborate with them in a distributed manner to render expedited Transformer inference at the edge.
As illustrated in Fig. \ref{fig:intro}, we can utilize the distributed computing resources in a smart home (with tablet, smart speaker, and television) to accelerate the Transformer-based (such as Bert \cite{devlin2018bert} and GPT \cite{radford2019language-gpt2}) voice assistant. Nevertheless, this paradigm brings several key challenges: (1) how to parallelize the single-shot Transformer inference workload among multiple edge devices, (2) how to decide the workload partitioning strategy tailored to the resource budget of heterogeneous edge devices, (3) how to reduce distributed inference latency under bandwidth-limited edge environments.

\begin{figure}[!t]
    \setlength{\abovecaptionskip}{-0.1cm}
    \centering
    \includegraphics[width=0.9\linewidth]{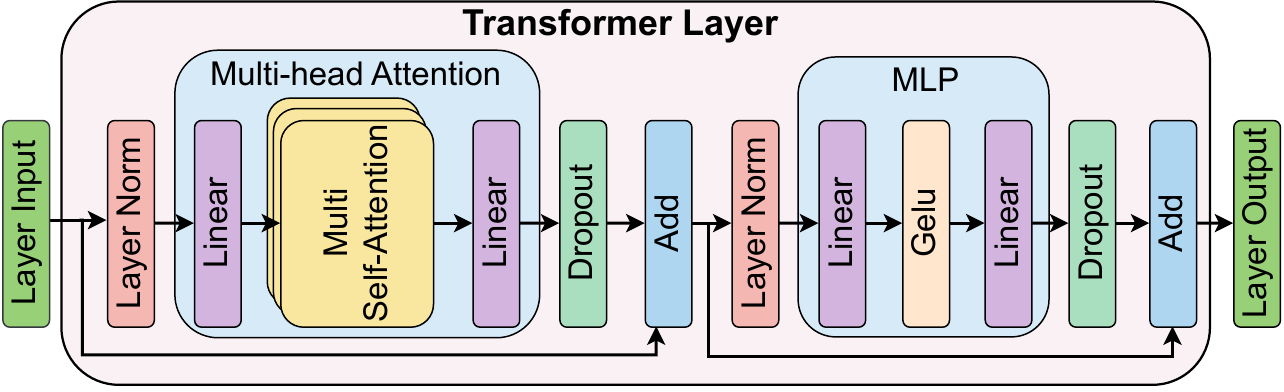}
    \caption{The architecture of a Transformer layer.}
    \label{fig:transformer}
    \vspace{-15pt}
\end{figure}

To address these challenges, we propose Galaxy, a collaborative edge AI system that breaks the resource walls across heterogeneous edge devices for low-latency Transformer inference to enable real-time in-situ edge intelligent services. Galaxy’s contribution goes beyond merely leveraging distributed edge devices for deploying Transformer inference, instead it addresses the above challenges on three levels. First, to orchestrate heterogeneous assisted devices in maximal resource utilization to facilitate collaborative inference, a novel hybrid model parallelism (HMP) that incorporates the best of both Tensor Parallelism (TP) and Sequence Parallelism (SP) is introduced as a novel parallel architecture to manage the distributed inference workflow. Second, to maximize resource utilization of HMP among edge devices, a workload planning algorithm that comprehensively accounts for both devices' resource heterogeneity and memory budget is equipped. Third, to achieve low-latency collaborative inference in bandwidth-limited edge environments, we meticulously decouple the tight data dependency between consecutive computation and communication operations by decomposing them into fine-grained tiles, thus enabling efficient overlapping for synchronization. Extensive evaluations on practical testbeds show that Galaxy achieves up to $2.5\times$ speed-up over the state-of-the-art collaborative inference approaches. A 4-way parallel inference with Galaxy can achieve $86\%$ scaling efficiency compared to the single device case. To the best of our knowledge, Galaxy is the first work to apply the hybrid model parallelism to edge collaborative Transformer inference scenarios.

In summary, this paper makes the following contributions. 
\begin{itemize}[leftmargin=*]
    \item Through extensive measurement studies on on-device and parallel inference methods, we introduce a novel HMP architecture to collaborate with trusted edge devices for in-situ single-shot Transformer inference acceleration.
    \item We devise a heterogeneity and memory-budget aware workload planning algorithm to facilitate resource-efficient edge collaborative inference.
    \item We propose a tile-based fine-grained optimization that leverages the concept of communication and computation overlapping to mitigate the synchronization overhead.
    \item We implement Galaxy and evaluate it in realistic edge testbeds. Experimental results show up to $2.5\times$ latency reduction over the state-of-the-art methods. 
\end{itemize}

\begin{table}[t!]
\caption{Inference Latency and Mem. Footprint of Transformer models}
\label{tab:execution}
\centering
\begin{tabular}{|c|c|c|c|c|c|}
\hline
Model & DistilBert & Bert-L & GPT2-L & OPT-L & OPT-XL \\ \hline \hline
Nano-M & 0.37s & 2.43s & OOM & OOM & OOM \\ \hline
Nvidia A100 & 5ms & 20ms & 29ms & 27ms & 38ms \\ \hline \hline
\begin{tabular}[c]{@{}c@{}}Memory\\ Footprint\end{tabular} & 130MB & 680MB & 1.6GB & 2.6GB & 5.4GB \\ \hline
\end{tabular}
% \vspace{-15pt}
\end{table}

\begin{figure}[t!]
    \setlength{\abovecaptionskip}{-0.1cm}
    \centering
    \includegraphics[width=0.9\linewidth]{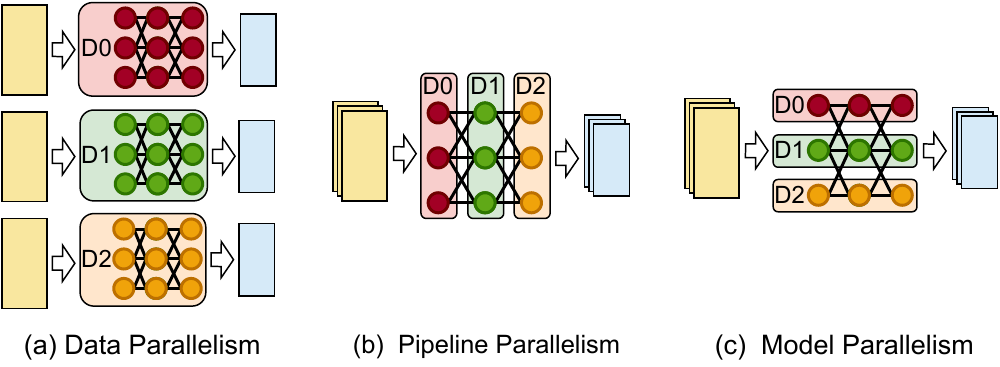}
    \caption{Different parallelism plans of collaborative Transformer inference.}
    \label{fig:paral}
    \vspace{-15pt}
\end{figure}

\begin{figure*}[!t]
    \setlength{\abovecaptionskip}{-0.1cm}
    \centering
    \includegraphics[width=0.95\linewidth]{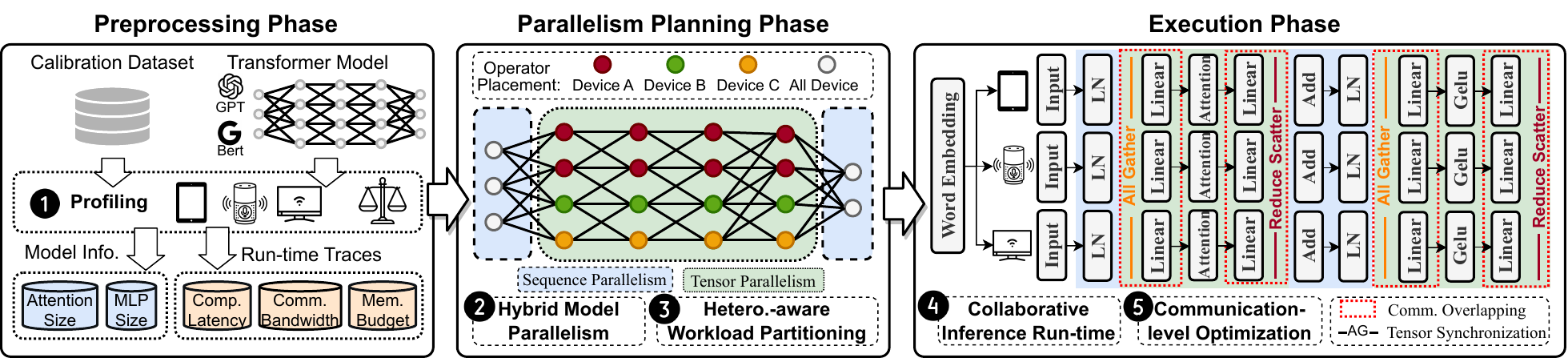}
    \caption{Galaxy system overview.}
    \label{fig:overview}
    \vspace{-15pt}
\end{figure*}

\section{Background and Motivation}

\subsection{Transformer-Based Model}
Current language-related applications trend towards using Transformer-based models, which are composed of stacks of Transformer layers, due to their superior accuracy. 
The original Transformer formulation \cite{vaswani2017attention} comprises both an \textit{Encoder} and a \textit{Decoder}. In this paper, we focus on the recent language models like Bert \cite{devlin2018bert} and GPT-2 \cite{radford2019language-gpt2}, which use only the \textit{Encoder} or \textit{Decoder} components. Fig. \ref{fig:transformer} shows the model architecture of the Transformer layer we consider in this paper.

In a Transformer layer, the primary components are the Multi-head Attention (MHA) block and the Multilayer Perceptron (MLP) block. These components are connected by element-wise operations such as Dropout, Residual Addition, and Layer Norm. In MHA block, the first linear layer generates query (Q), key (K), and value (V) matrices for each attention head. Each head conducts self-attention independently, and their outputs are concatenated and further processed through a final linear layer to obtain the output.
MLP block involves two linear operations which increase the hidden size from $h$ to $4h$ and then reduce it back to $h$.

\subsection{Transformer Inference on Resource-Limited Edge Devices}
\label{sec:motivation-edge}

In-situ inference can leverage idle resources in edge environments while fully preserving users' data privacy, making it a widely utilized paradigm in privacy-sensitive edge applications \cite{bhardwaj2022ekya,zeng2020coedge}. However, the resource-intensive nature of Transformer inference presents significant challenges for resource-limited edge devices \cite{aminabadi2022deepspeed, lin2022device}. 
We conduct experiments to analyze how limited computation resources affect on-device Transformer inference. The experimental setup is described in \S \ref{subsec:exp-setup}, and the results are presented in Table \ref{tab:execution}. Specifically, we perform on-device inference for five typical Transformer-based models on off-the-shelf edge devices and the Nvidia GPU platform using an input sequence of length 30. We observe that the inference latency exhibits a huge gap between A100 and Nano-M, e.g., $121\times$ slowdown for Jetson Nano when comparing with A100 on Bert-L. Memory budget is another critical factor in Transformer inference. GPT2-L in half-precision floating-point format incurs a 1.6GB memory footprint during inference, exceeding the 1.5GB budget of a single Nano-M.
To mitigate resource constraints, we leverage our observation that edge environments often consist of multiple trusted edge devices in physical proximity. This enables mutually-trustworthy computation resource sharing among these edge devices \cite{ye2022eco,zeng2020coedge}.

\subsection{Collaborative Transformer Inference with Multiple Devices}
In collaborative Transformers inference across edge devices, the key question is the choice of parallelism strategy. We illustrate different parallelism plans in Fig. \ref{fig:paral}.

\begin{figure*}[!t]
    \setlength{\abovecaptionskip}{-0.1cm}
    \centering
    \includegraphics[width=0.91\linewidth]{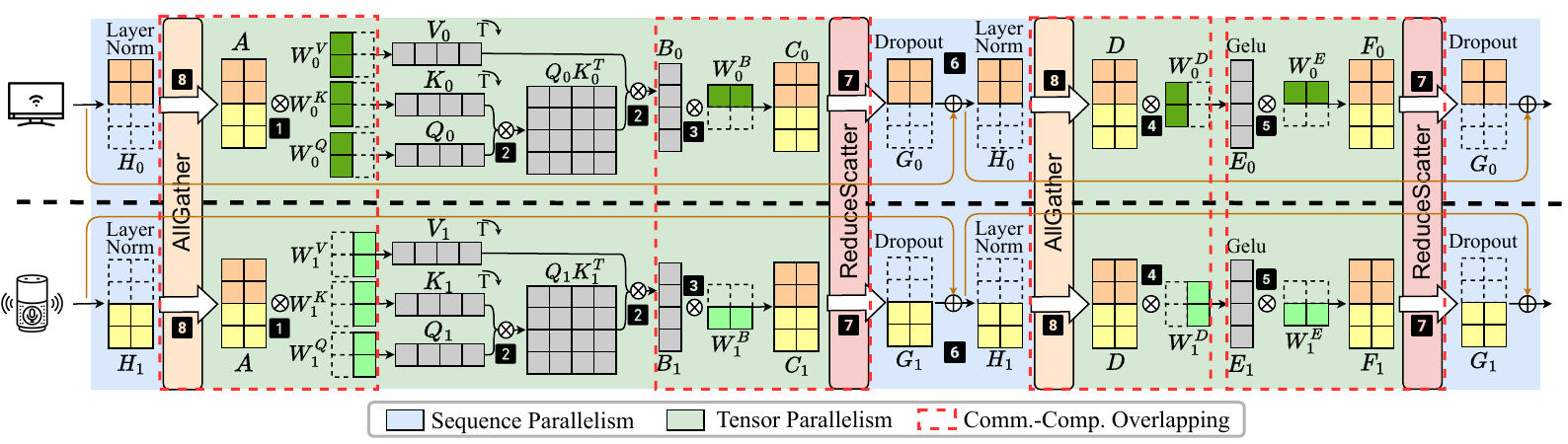}
    \caption{Hybrid model parallelism in matrix form.}
    \label{fig:hmp}
    \vspace{-15pt}
\end{figure*}

\subsubsection{Data and Pipeline Parallelism}
Data Parallelism (DP) and Pipeline Parallelism (PP) are the common way to execute Transformer-based model in parallel \cite{li2014communication,huang2019gpipe,narayanan2019pipedream}. 
DP partitions workloads along the sample dimension, allowing each device to perform inferences independently.
In edge intelligence services, where single-shot inference requests are frequently raised (e.g., sending a single piece of voice command to a smart assistant), DP is not applicable due to the absence of data batches.
PP horizontally partitions the model into consecutive stages along layer dimension, with each stage mapped to a distinct device. 
However, in the case of single-shot inference, PP still falls short in leveraging multiple edge devices concurrently, as the inter-stage data dependencies force each device to await completion of the preceding one.

\subsubsection{Model Parallelism}
Model Parallelism (MP) is a parallel computing paradigm that horizontally partitions the operations within a model layer, facilitating concurrent execution of single-shot inference. The most common techniques of model parallelism applied to Transformer models are Tensor Parallelism (TP) \cite{shoeybi2019megatron, aminabadi2022deepspeed} and Sequence Parallelism (SP) \cite{li2021sequence}.
TP partitions model weights across devices, each hosting a subset of parameters, yet it fails to parallelize some element-wise operations between MHA and MLP block. In contrast, SP segments the input along the sequence dimension, facilitating parallelism for all operations, but requires each device to store the entire model parameters.
Due to intra-layer data dependencies, synchronization points are inserted during MP to ensure consistency between collaborative and local inference results. 
However, these synchronization points introduce significant communication latency, potentially becoming a bottleneck in inference performance, especially in bandwidth-limited edge environments.

Summarizing the above analysis motivates our design of a hybrid model parallelism architecture that incorporates the best of both TP and SP, with a communication optimization approach to mitigate synchronization overhead.

\section{Galaxy Design}
\subsection{Galaxy Workflow} 
Our system design aims to concurrently utilize multiple heterogeneous edge devices to achieve low-latency in-situ Transformer inference.
Fig. \ref{fig:overview} illustrates the workflow of our proposed Galaxy, which features three primary phases: \textit{Preprocessing Phase}, \textit{Parallelism Planning Phase} and \textit{Execution Phase}. \textit{Preprocessing Phase} is an offline procedure that runs once before deployment. \textit{Galaxy Profiler} performs an inference process using calibration data as input on the physical edge devices to record the run-time traces necessary for parallelism planning (step \circled{1}). In parallelism planning phase, Galaxy adopts a novel hybrid model parallelism (HMP) architecture that incorporates both TP and SP to orchestrate distributed edge devices (step \circled{2}). \textit{Galaxy Planner} takes profiling results from \textit{Galaxy Profiler} as input to generate a parallelism planning configuration (step \circled{3}). This configuration comprehensively considers both resource heterogeneity and memory budget, and is subsequently applied to target models and edge devices in \textit{Execution Phase} for efficient edge collaborative inference (step \circled{4}). Distributed inference inevitably involves tensor synchronization operations. Galaxy incorporates a tile-based fine-grained communication optimization to mitigate the performance degradation brought by additional communication overhead (step \circled{5}).
With the above modules, Galaxy focuses on the following design goals: 
\begin{itemize}[leftmargin=*]
    \item A HMP architecture for low-latency single-shot Transformer inference across multiple edge devices (\S \ref{sec:model-parallelism}).
    \item A judicious parallel planner that comprehensively considers the device heterogeneity and memory budget, aiming at distributing workload in a load-balanced manner to fully exploit computing resources of edge devices (\S \ref{sec:planning}).
    \item A tile-based fine-grained communication optimization decouples the tight dependency between consecutive computation and communication operations, enabling efficient overlapping between them (\S \ref{sec:communication}).
\end{itemize}

\subsection{Hybrid Model Parallelism}
\label{sec:model-parallelism} 
Galaxy incorporates an innovative HMP architecture that facilitates efficient parallel Transformer inference within edge environments.
In this section, we will elaborate on our HMP architecture, using an example of collaborative inference conducted across two edge devices.
As illustrated in Fig. \ref{fig:hmp}, TP and SP alternate throughout a Transformer layer. Specifically, TP is applied to the MHA block and MLP block while SP is applied to operations connecting the MHA and the MLP blocks, namely connective blocks.

\subsubsection{Tensor Parallelism on MHA Block} The aim of designing an efficient TP approach is to reduce the data dependencies among operators split across various devices, thereby reducing the frequency of tensor synchronization \cite{shoeybi2019megatron, narayanan2021efficient}.
As illustrated in Fig. \ref{fig:hmp}, the first block applied with TP is the MHA block.
We exploit the inherent parallelism advantage of MHA: the computation of multiple attention heads is entirely independent. This head-level dependency allows us to split the operations of each attention head across edge devices without any tensor synchronization during the execution of Multi Self-Attention operations.
With this in mind, we partition the weight matrices associated with key ($W^{K}$), query ($W^{Q}$), and value ($W^{V}$) along their head dimension. The initial General Matrix Multiply (GEMM) is distributed to distinct devices and parallelized along head dimension (\squared{1}). Subsequently, Self-Attention corresponding to each attention head is carried out locally on each respective device (\squared{2}). The final GEMM from the output linear layer is parallelized along its row dimension, ensuring alignment with the initial GEMM's head-wise partition (\squared{3}). The operations on device $i$ ($i \in \{0,1\}$) can be formulated as follows ([·$|$·]is the concat operation):
\begin{equation}
\begin{aligned}
&[Q_i|K_i|V_i] = [W_{i}^Q|W_{i}^K|W_{i}^V]\cdot A, \\
&B_i = \operatorname{Self-Attention}(Q_i,K_i,V_i), \\
&C_i = W_i^{B}B_i.
\end{aligned}
\end{equation}

\subsubsection{Tensor Parallelism on MLP Block}
As illustrated in Fig. \ref{fig:hmp}, the second block applied with TP is the MLP block, which comprises two consecutive GEMMs. To obviate tensor synchronization between the first and second GEMM operations, we leverage the concept of matrix tiling to remove data dependencies. We partition the weight matrix of the first GEMM along its column dimension (\squared{4}), and partition the second GEMM along its row to align with the column-wise 
partition of the first GEMM (\squared{5}). 
The second GEMM can directly take the output of the first GEMM as input without a synchronization point. 
The operations on device $i$ ($i \in \{0,1\}$) can be formulated as follows:
\vspace{-0.1cm}
\begin{equation}
\begin{aligned}
&E_i = \operatorname{GELU}(W_i^DD), \\
&F_i = W_i^{E}E_i.
\end{aligned}
\end{equation}

\subsubsection{Sequence Parallelism on Connective Block}
TP expedites the most computationally intensive parts of each Transformer layer while leaving the Dropout, Residual Addition and Layer Norm connecting the MHA block and the MLP block untouched (\squared{6}).
Although these operations are element-wise and entail no intensive matrix multiplication, they require a considerable amount of memory access, thus also yielding a non-negligible execution latency. 
We notice that these element-wise operations are independent along the sequence dimension which allows us to parallelize them by partitioning the input sequence. The operations on device $i$ ($i \in \{0,1\}$) can be formulated as follows:
\vspace{-0.1cm}
\begin{equation}
H_i = \operatorname{Layernorm}(\operatorname{ResidualAdd}(\operatorname{Dropout}(G_i))).
\end{equation}

\subsubsection{Tensor Synchronization Points}
To ensure that the inference results from our HMP align with the local inference results, a synchronization point is required at the end of each TP and SP block, as illustrated in Fig. \ref{fig:hmp}. 

Towards the completion of TP blocks, a \textit{ReduceSum} operation is required to aggregate the computation results across multiple devices ($G\leftarrow C_0+C_1$ and $G\leftarrow F_0+F_1$). Subsequently, the aggregated results are partitioned along the sequence dimension and scattered across various edge devices for SP ($[G_0|G_1]\leftarrow G$).
These two operations can be efficiently combined and implemented using a single \textit{ReduceScatter} operation (\squared{7}).
% These two operations can conveniently be combined and implement with a single \textit{ReduceScatter} operation (\squared{7}).
Towards the completion of SP blocks, each device retains only a segment of the input sequences. It is essential to gather all these fragments, concatenate them, and distribute them across all devices for subsequent TP ($A\leftarrow [H_0|H_1]$ and $D\leftarrow [H_0|H_1]$). Consequently, we perform an \textit{AllGather} communication primitive at the end of each SP block (\squared{8}).

\subsubsection{Merits of Hybrid Model Parallelism Architecture} Employing the HMP architecture presents numerous advantages over straight TP or SP architecture. \textbf{Compared to TP}: (1) the HMP architecture eliminates redundant computations in the connective blocks, which fully exploits the parallel potential of Transformer layers. (2) HMP does not introduce additional communication overhead. At first glance, state-of-the-art TP \cite{shoeybi2019megatron} requires two \textit{AllReduce}, while the HMP requires two \textit{ReduceScatter} and two \textit{AllGather} operations per Transformer layer. However, in the implementation of communication primitives, the communication volume of a single \textit{Ring-AllReduce} operation equates to a \textit{Ring-ReduceScatter} followed by a \textit{Ring-AllGather} \cite{sergeev2018horovod}. (3) HMP architecture split a larger \textit{AllReduce} operation into two smaller primitives, \textit{ReduceScatter} and \textit{AllGather}, which greatly facilitates our tiled-based communication overlapping proposed in \S \ref{sec:communication}.
\textbf{Compared to SP}: SP partitions the input tensor along sequence dimension without partitioning the weight matrices. This paradigm requires each device to accommodate a holistic copy of the global model. HMP mitigates this issue by distributing model parameters across devices, thereby breaking the memory wall of individual devices and achieving memory resource scalability.

\subsection{Heterogeneity and Memory Aware Workload Planning}
\label{sec:planning}
Fig. \ref{fig:hmp} shows that a synchronization point is required after each TP or SP block completion. The initiation of these synchronization points is bound by the completion time of the slowest device (straggler). Such straggler can starve other faster devices, resulting in resource under-utilization. Given the inherent heterogeneity in computing capacities of devices, particularly notable in edge environments, adopting a heterogeneity-aware workload planning is essential to distribute the workload in a balanced manner.
Furthermore, inference on Transformer-based models necessitates considerable memory. In practical deployment, an out-of-memory (OOM) issue is a game-stopper for inference, which poses substantial challenges for edge devices that usually operate within tight memory limitations. Consequently, our workload planning should also comprehensively consider each device's memory budget to prevent overconsumption of available memory.

\subsubsection{Optimization Target Formulation}
As elaborated in \S \ref{sec:model-parallelism}, our HMP architecture allocates workload by partitioning along three distinct dimensions: the head dimension for the MHA block, the row dimension of the weight matrix for the MLP block, and the sequence dimension of the input tensor for the connective block. Our workload planning focuses on determining the partition configuration for each of these blocks, namely: the MHA blocks partition $\mathcal{A}=\{a_0, a_1,...,a_{\mathcal{D}-1}\}$, the MLP blocks partition $\mathcal{B}=\{b_0, b_1,...,b_{\mathcal{D}-1}\}$, and the connective blocks partition $\mathcal{S}=\{s_0, s_1,...,s_{\mathcal{D}-1}\}$, where $\mathcal{D}$ is the number of edge devices.
We introduce the notation $L(\text{MHA},\mathcal{A}_d, d)$, $L(\text{MLP},\mathcal{B}_d, d)$, and $L(\text{CON},\mathcal{S}_d, d)$ to represent the execution latency of the MHA block, the MLP block, and the connective block on device $d$, respectively, each given their partition configurations $\mathcal{A}_d$, $\mathcal{B}_d$, and $\mathcal{S}_d$.
The execution time $\mathcal{L}$ for each TP or SP block is determined by the straggler:
\begin{equation}
\begin{aligned}
& \mathcal{L}(\text{MHA}, \mathcal{A}) = \max_{d \in \{0,1,...,\mathcal{D}-1\}} L(\text{MHA},\mathcal{A}_d,d), \\
& \mathcal{L}(\text{MLP}, \mathcal{B}) = \max_{d \in \{0,1,...,\mathcal{D}-1\}} L(\text{MLP},\mathcal{B}_d,d), \\
& \mathcal{L}(\text{CON}, \mathcal{S}) = \max_{d \in \{0,1,...,\mathcal{D}-1\}} L(\text{CON},\mathcal{S}_d,d). \\
\end{aligned}
\end{equation}
  
Beyond minimizing the execution latency, our strategy also requires to prevent OOM errors during inference. The overwhelming memory footprint in deploying Transformer-based models stems from the substantial weight matrices housed within the MHA and MLP blocks.
Therefore, our workload planning judiciously partitions the MHA and MLP blocks, allowing the memory demands of the model to be collaboratively handled by multiple devices.
We denote $M_{att}$ and $M_{mlp}$ as the memory footprint of loading one MHA block and one MLP block, respectively. $\text{Budget}_d$ denotes the memory budget allocated to device $d$, and $l$ represents the total number of Transformer layers within the model.
Putting them together, the optimization objective for minimizing the latency under memory constraints is as follows:
\vspace{-0.2cm}
\begin{equation}
\begin{aligned}
\min _{\mathcal{A}, \mathcal{B}, \mathcal{S}} \biggl(\mathcal{L}(\text{MHA}, \mathcal{A})+\mathcal{L}(\text{MLP}, \mathcal{B})+  \mathcal{L}(\text{CON}, \mathcal{S})\biggr), \\ 
\text{s.t.} \quad l \cdot (M_{att} \cdot \frac{a_d}{\sum{\mathcal{A}}}+M_{mlp} \cdot \frac{b_d}{\sum{\mathcal{B}}}) <\operatorname{Budget}_d, \\ 
\text{where} \ d \in\{0,1, \cdots \mathcal{D}-1\}.\\
\end{aligned}
\end{equation}

To facilitate our workload planning algorithm, we employ \textit{Galaxy Profiler}, which conducts an inference process using calibration dataset as input on the physical edge devices to record the run-time profile necessary for parallelism planning. The profiler meticulously captures the computation latency under a variety of partition configurations, for both TP and SP blocks. Simultaneously, \textit{Galaxy Profiler} also records the model information, involving the number of parameters contained within the MHA and MLP blocks.

\begin{algorithm}[t]
\normalem
\setlength{\textfloatsep}{0.5cm}
\setlength{\floatsep}{0.5cm}
\setlength{\intextsep}{-1em} 
\caption{Heterogeneity and Memory Aware Workload Planning}\label{alg:scheduling}
\label{alg:workload-planning}
\KwIn{Profiling results of models and devices. $\mathcal{V}$: The list of computing capacity of devices.}
\KwOut{$\mathcal{A},\mathcal{B}$: Partition configurations of MHA and MLP block.}

\SetKwRepeat{Do}{do}{while}
\SetKwFunction{BP}{BalacedPartition}
\SetKwFunction{MAB}{MemoryAwareBalancing}
\SetKw{Return}{Return}
\SetKwProg{Ft}{Function}{:}{}

\Ft{\BP{$T, \mathcal{V}$}}{
    Initialize partition configuration $C$\;
    Workload $\leftarrow$ Total workload in block $T$\;
    \ForEach{$d \in \{0, 1, 2,...,\mathcal{D}-1\}$}{
        $C_d \leftarrow (\mathcal{V}_d / \sum\mathcal{V}) \cdot \text{Workload}$\;
    }
    \textbf{Return $C$}; 
}

$\mathcal{A} \leftarrow $ \BP{$\text{MHA}$,$\mathcal{V}$}\; $\mathcal{B} \leftarrow $ \BP{$\text{MLP}$,$\mathcal{V}$}\;

\Ft{\MAB{$T, C, \mathcal{V}$,$\mathcal{L}$}}{
    $OOM\_Devices \leftarrow$ Out-of-memory devices under partition configuration $C$ in $\mathcal{L}$\;
    $Free\_Devices \leftarrow$ Devices retaining available memory under partition config. $C$ in $\mathcal{L}$\;
    \uIf{$OOM\_Devices = \emptyset$}{
    \textbf{Return $C$}\; 
    % $\quad \quad \;\;\triangleright$ Success and Return\\  
    }
    \ForEach{$o \in OOM\_Devices$}{
        Waiting\_Shift $\leftarrow$ Overflowing workload on device $o$\;
        \ForEach{$f \in Free\_Devices$}{
            % max\_shift $\leftarrow$ Maximum number of workload can shift from $o$ to $f$ without OOM\;
            Shift $(\mathcal{V}_f/\sum_{i\in Free\_Devices} \mathcal{V}_i) \cdot \text{Waiting\_Shift}$ workload from $o$ to $f$\;
        }
        Remove device $o$ from $\mathcal{L}$\;
        % Update \underline{$OOM\_Devices$} and \underline{$Free\_Devices$} lists\;
    }
    \MAB{$T,C,\mathcal{V}$,$\mathcal{L}$}\;
    % $\quad \quad \;\;\triangleright$ Tail recursion\\  
}

$\mathcal{L} \leftarrow$ $[0,1,..,\mathcal{D}-1]$; $\quad \;\;\triangleright$ List of all devices \\
$\mathcal{B} \leftarrow$ \MAB{\text{MLP}, $\mathcal{B}, \mathcal{V}$,$\mathcal{L}$}\; 
% $\quad \quad \;\;\triangleright$ Re-distribute MLP weights and update $\mathcal{B}$\\
$\mathcal{A} \leftarrow$ \MAB{\text{MHA}, $\mathcal{A}, \mathcal{V}$,$\mathcal{L}$}\; 
% $\quad \quad \;\;\triangleright$ Re-distribute Attention heads and update $\mathcal{A}$\\
\uIf{Out-of-memory devices still exist}{
    \textbf{Exit with Fail}\; 
    % $\quad \quad \;\;\triangleright$ Target model fails to fit into devices' memory\\  
    }

% \BP{}; $\quad \quad \;\;\triangleright$ Phase 2
\end{algorithm}

\subsubsection{Workload Planning Algorithm}
A straw-man approach to address the above constrained optimization problem would involve an exhaustive search of all potential partitioning combinations, subsequently selecting the optimal solution that satisfies the memory constraints. However, this method suffers from an exponential complexity, rendering it infeasible for large-scale Transformer models.
% of $O((\sum \mathcal{A}\sum \mathcal{B} \sum \mathcal{S})^{\mathcal{D}-1})$, 

The connective block's execution time hinges primarily on memory access volume rather than the SoC's computing capabilities, where we adopt a strategy of \textit{equal partition} for SP planning.
Equal partition preserves uniform communication volume across all devices during tensor synchronizations, laying a conducive foundation for our tile-based communication overlapping in \S \ref{sec:communication}.
Towards TP, we can achieve optimal partitioning of blocks with workload distribution proportional to each device's computing capacity, disregarding the memory budget. This \textit{proportional partition} ensures that all devices complete their tasks almost simultaneously, effectively mitigating potential delays that might lead to suboptimal resource utilization. 
With these insights, we devise a two-step heuristic algorithm, outlined in Algorithm \ref{alg:workload-planning}. In the first step, the algorithm disregards the memory constraints of the devices and distributes the workload commensurate with their computing capacities, thereby ensuring a balanced workload (lines 1-8). Subsequently, building on this initial distribution, the second step fine-tunes the workload allocation. It redistributes excess workloads from devices that surpass their memory budgets to those with spare memory capacity. (lines 9-19). Considering that the granularity of partitioning for MHA block (head dimension) is typically coarser than that of MLP block (column dimension), we first redistribute the workload for MLP block (line 21), followed by MHA block (line 22). If OOM errors persist despite workload redistribution, this indicates that the edge devices involved in collaborative inference are not capable of accommodating the target model, thus resulting in the algorithm's failure (lines 23-24).
We define a device’s computing capacity $\mathcal{V}_d$ as the inverse of the total time required to execute a MHA block and a MLP block on device $d$. 
\begin{equation}
\begin{aligned}
\mathcal{V}_d = \biggl( L(\text{MHA},\sum\mathcal{A},d)&+ L(\text{MLP},\sum\mathcal{B},d) \biggr)^{-1}.
\end{aligned}
\end{equation}
The workload planning is an offline procedure that runs once before deployment. The time complexity for Algorithm \ref{alg:workload-planning} exhibits a upper bound of $O(\mathcal{D}^3)$. In our experiment, the running time is under one second on a domestic desktop for 4 heterogeneous edge devices.

\subsection{Tile-based Communication Optimization}
\label{sec:communication}

In contrast to stable, high-bandwidth networks in datacenters, edge environments frequently grapple with inconsistent, bandwidth-limited connections. This amplifies synchronization latency during the collaborative inference, serving as a significant bottleneck of global system performance. 
Overlapping communication and computation is an effective optimization strategy. However, its implementation becomes intricate in the Transformer inference due to the strict data dependencies between communication and computation. To address this, Galaxy introduces a tile-based approach to effectively decouples their dependency to achieve a fine-grained overlapping.
We observe from Fig. \ref{fig:hmp} that each TP block starts and ends with GEMM operations. 
We design to overlap these GEMM operations with the AllGather and ReduceScatter operations when entering and exiting the TP blocks. To illustrate this, the following section provides an example of collaborative inference across three devices, demonstrating how to overlap GEMMs with synchronization points before and after the MLP blocks (also applicable to the MHA blocks).

\begin{figure}[t!]
    \setlength{\abovecaptionskip}{-0.1cm}
    \centering
    \includegraphics[width=\linewidth]{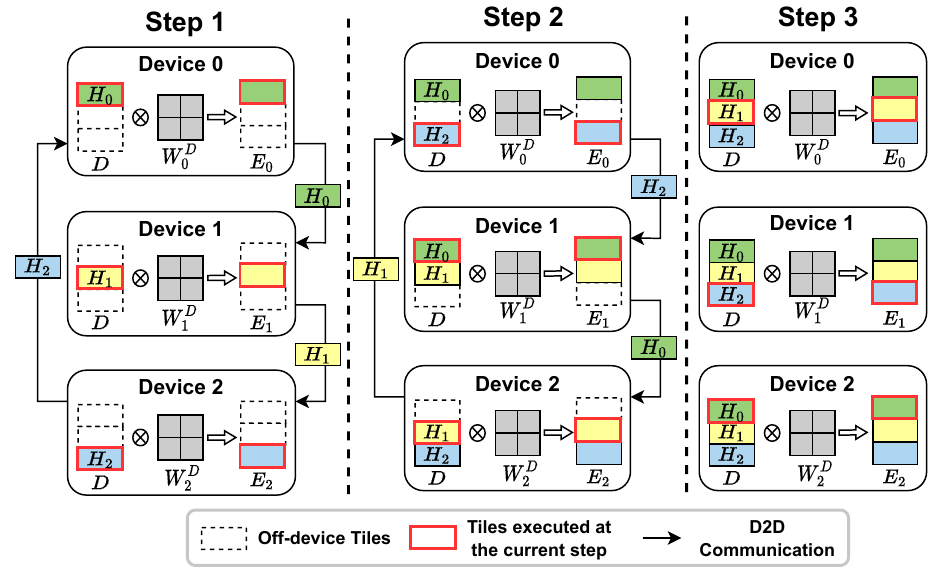}
    \caption{Ring-AllGather overlapping.}
    \label{fig:all-gather}
    \vspace{-15pt}
\end{figure}

\subsubsection{AllGather Overlapping}
As illustrated in Fig. \ref{fig:hmp}, a strict data dependency exists between the AllGather and the initial matrix multiply (GEMM1) in MLP block. Specifically, GEMM1 on device $i$ ($i \in \{0,1,2\}$) can only commence after the AllGather has finished aggregating all sub-sequences:
\begin{equation}
    D = \operatorname{AllGather}(H_0,H_1,H_2), E_i = \operatorname{GEMM1}(D,W_i^D).
\end{equation}
To decouple the strict dependency between AllGather and GEMM1, we leverage matrix tiling to decompose GEMM1. We discover that the direct calculation of GEMM1 can be equivalently achieved by segmenting matrix D horizontally into tiles, executing the GEMM1 independently on each tile, and subsequently concatenating the results.
\begin{equation}
E_i=
\left[
\begin{array}{c}
H_0\cdot W_i^D \\
\hline
H_1\cdot W_i^D \\
\hline
H_2\cdot W_i^D \\
\end{array}
\right]
=
\left[
\begin{array}{c}
H_0 \\
\hline
H_1 \\
\hline
H_2 \\
\end{array}
\right]
\cdot
W_i^D
= D \cdot W_i^D.
\end{equation}

We employ a \textit{Ring-AllGather} implementation and integrate it with the above matrix tiling approach to overlap communication and computation. An example of an overlapping process involving three collaborative devices is illustrated in Fig. \ref{fig:all-gather}.
In the context of a tile-based overlapping process that incorporates $\mathcal{D}$ devices, typically $\mathcal{D}$ steps are required (three steps in this case). 
We define $(i+1)$$\%3$ and $(i-1)$$\%3$ represent the index of succeeding and preceding device of device $i$ within a 3-device ring topology. 
\textbf{Step 1}: Device $i$ performs GEMM operation between on-device tile $H_i$ and $W_i^D$, and concurrently dispatches $H_i$ to the succeeding device. In parallel, Device $i$ receives and stores the tile $H_{(i-1)\%3}$ transmitted from its preceding device.
\textbf{Step 2}: Device $i$ performs GEMM operation on tile $H_{(i-1)\%3}$ and concurrently dispatches it to the succeeding device. In parallel, Device $i$ receives the tile $H_{(i-2)\%3}$ transmitted from its preceding device. 
\textbf{Step 3}: Device $i$ executes the GEMM operation on the tile $H_{(i-2)\%3}$. Notably, the final step does not necessitate any communication. The outcomes of the three GEMM operations are concatenated along the sequence dimension, yielding the final result $E_i$.

\begin{figure}[t!]
    \setlength{\abovecaptionskip}{-0.1cm}
    \centering
    \includegraphics[width=1.05\linewidth]{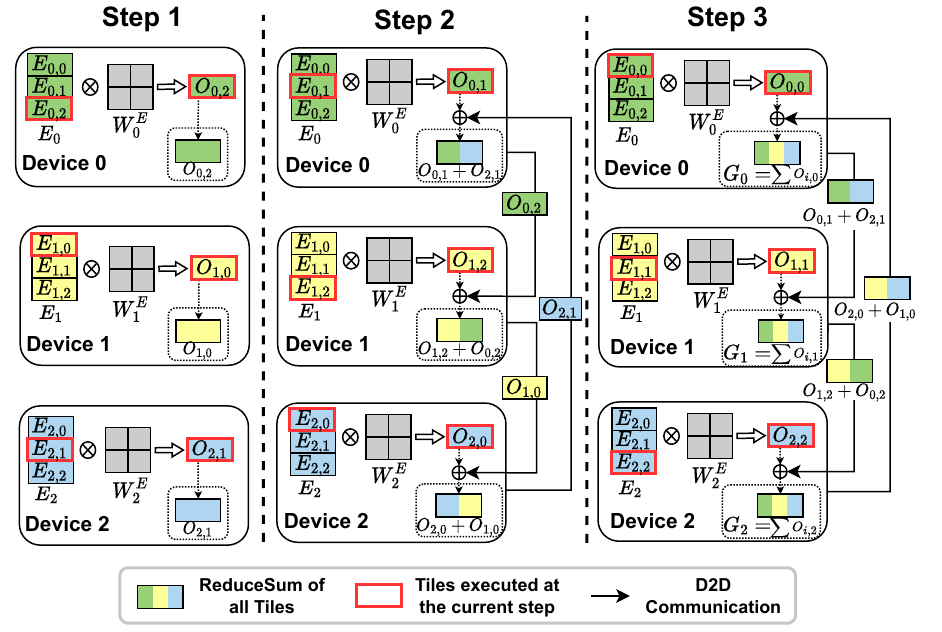}
    \caption{Ring-ReduceScatter overlapping.}
    \label{fig:reduce-scatter}
    \vspace{-15pt}
\end{figure}

\subsubsection{ReduceScatter Overlapping}
As illustrated in Fig. \ref{fig:hmp}, a strict data dependency exists between the final matrix multiplication (GEMM2) in the MLP block and the ReduceScatter operation ($i \in \{0,1,2\}$):
\begin{equation}
    F_i = \operatorname{GEMM2}(E_i, W_i^E), G_i = \operatorname{ReduceScatter}(F_0,F_1,F_2).
\end{equation}
To decouple the strict dependency between ReduceScatter and GEMM2,
we mirroring the tiling approach used with the AllGather. We split the matrix $E_i$ into three equally-sized tiles $E_{i,r}$ ($r \in \{0,1,2\}$) along the row dimension (aligns with the partition configuration of connective block) and compute GEMM2 independently for each tile (Eq.\ref{equ:Oir}). To obtain the final result $G_r$, an additional ReduceSum operation across all devices is necessary (Eq.\ref{equ:Gr}).

% ReduceScatter incorporates an additional ReduceSum operation to obtain the final result $G_r$:

\begin{equation}
\label{equ:Oir}
\left[
\begin{array}{c}
O_{i,0} \\
\hline
O_{i,1} \\
\hline
O_{i,2} \\
\end{array}
\right]
=
\left[
\begin{array}{c}
E_{i,0} \cdot W_i^E \\
\hline
E_{i,1} \cdot W_i^E \\
\hline
E_{i,2} \cdot W_i^E \\
\end{array}
\right]
=
\left[
\begin{array}{c}
E_{i,0} \\
\hline
E_{i,1} \\
\hline
E_{i,2} \\
\end{array}
\right] \cdot W_i^E
=E_i\cdot W_i^E,
\end{equation}
\begin{equation}
\label{equ:Gr}
G_r = \sum_i O_{i,r}.
\end{equation}

Similar to AllGather, we employ a \textit{Ring-ReduceScatter} implementation coupled with matrix tiling to achieve communication and computation overlapping. As illustrated in Fig. \ref{fig:all-gather}, the process of ReduceScatter overlapping also involves three steps. 
\textbf{Step 1}: Device $i$ performs GEMM operation between tile $E_{i,(i+2)\%3}$ and $W_i^E$, yielding the result $O_{i,(i+2)\%3}$.
\textbf{Step 2}: Device $i$ perform GEMM operation on tile $E_{i,(i+1)\%3}$ and yield the result $O_{i,(i+1)\%3}$. In parallel, device $i$ forwards the GEMM result in step 1 to the subsequent device. Upon receiving the tile from the preceding device, Device $i$ conducts a ReduceSum operation between it and $O_{i,(i+1)\%3}$.
\textbf{Step 3}: Device $i$ perform GEMM operation on tile $E_{i,i}$ and yield the result $O_{i,i}$. Device $i$ concurrently sends the result of ReduceSum in Step 2 to the subsequent device. A ReduceSum operation is performed between the tile received from the preceding device and $O_{i,i}$, yielding the final result $G_i$.

Our tile-based communication optimization seamlessly overlaps $\mathcal{D}-1$ rounds of ring communication with $\mathcal{D}$ rounds of GEMM operation, without imposing additional overhead or yielding results inconsistent with non-overlapping approaches.

\section{Implementation and Evaluation}
We have fully implemented the prototype system of Galaxy and baselines with $\sim$1500 LoC in Python and C/C++ atop Pytorch \cite{pytorch}.
% Galaxy is also extensible 
Galaxy's idea is also portable and can work well with other lightweight ML frameworks such as MNN \cite{jiang2020mnn} and TF-Lite \cite{tflite}. 
In this section, we evaluate the performance of Galaxy prototype for five different sizes of Transformer-based models on physical testbeds.

\subsection{Experimental Setup}
\label{subsec:exp-setup}
\textbf{Models and Datasets.} We evaluate Galaxy with five typical Transformer-based models ranging from 66 Million to 2.7 Billion parameters, as detailed in Table \ref{tab:model}. We extract a subset of samples where the average sequence length is 284 from QNLI corpus of popular GLUE datasets \cite{wang2018glue} for evaluation.

\textbf{Edge Environment Setup.} 
We evaluate Galaxy across a diverse range of realistic edge environments, incorporating both homogeneous and heterogeneous configurations of off-the-shelf edge devices (Jetson Nano \cite{jetson-nano}), as detailed in Table \ref{tab:pi} and \ref{tab:edge-env}.
In homogeneous environments, the memory budget for Nano-M is set at 1.5GB. In the heterogeneous environments, the memory budgets are set at 1.5GB for Nano-L, 1.2GB for Nano-M, and 0.7GB for Nano-S, respectively.
We limit usage to the onboard CPU to simulate resource-constrained edge scenarios. We will also demonstrate the effectiveness of Galaxy in GPU environments in \S \ref{sec:gpu}. We adjust the D2D bandwidth to simulate the diverse network conditions within realistic edge environments.

\textbf{Baseline Methods.} We compare Galaxy with both single-device method and state-of-the-art parallel methods:
\begin{itemize}[leftmargin=*]
    \item \textbf{Local Inference (Local)}: Inference models on a single device. We compare with it to analyze the scalability performance of Galaxy.
    \item \textbf{Megatron-LM (M-LM)}\cite{shoeybi2019megatron}: A state-of-the-art TP method splits the weight matrix in MHA and MLP blocks to parallelize the GEMM operators. An AllReduce synchronization is required after each MHA and MLP block.
    \item \textbf{Sequence Parallelism (SP)}\cite{li2021sequence}: A state-of-the-art SP method partitions the input along its sequence dimension and parallelizes inference across workers. Two AllGather synchronizations are required among each MHA block.
\end{itemize}

\begin{table}[t!]
\caption{Jetson Nano Specifications \cite{jetson-nano}}
\label{tab:pi}
\centering
\begin{tabular}{c|c|c}
\hline
Hardware & \multicolumn{2}{c}{Specifications} \\ \hline \hline
CPU      & \multicolumn{2}{c}{Quad Core ARM Cortex-A53 CPU}\\ \hline
GPU      & \multicolumn{2}{c}{128 Core Maxwell GPU} \\ \hline \hline
% Default Memory Budget      & \multicolumn{2}{c|}{1536MB} \\ \hline \hline
CPU Frequency Mode    & \begin{tabular}[c]{@{}c@{}}Nano-S\\ Nano-M\\ Nano-L\end{tabular} & \begin{tabular}[c]{@{}c@{}}403MHz\\ 825MHz\\ 1.47GHz\end{tabular} \\ \hline
\end{tabular}
\vspace{-10pt}
\end{table}

\begin{table}[t!]
\centering
\caption{Specifications of edge environments.}
\label{tab:edge-env}
\begin{tabular}{cc|cc}
\hline
ID & Homogeneous Edge Env. & ID & Heterogeneous Edge Env. \\ \hline \hline
A & 2 $\times$ Nano-M & D & Nano-L + Nano-M \\
B & 3 $\times$ Nano-M & E & Nano-L + Nano-S \\
C & 4 $\times$ Nano-M & F & Nano-L + Nano-M + Nano-S\\ \hline
\end{tabular}
\vspace{-15pt}
\end{table}

\subsection{Comparison to Baselines}
Table \ref{tab:model} summarizes the general performance results comparing Galaxy with state-of-the-art methods M-LM and SP. We conduct experiments on three different homogeneous edge environments with 125Mbps intra-cluster bandwidth.
We employ the average end-to-end inference latency as our performance metric.
The results indicate that owing to our HMP architecture and tile-based communication optimization, Galaxy outperforms baselines across various models and edge environments. 
Specifically, when comparing to M-LM, Galaxy achieves up to $1.46\times$ higher performance. With the increase in model size, the communication-to-computation ratio declines. This narrows the room for our communication optimization, correspondingly leading to a decrease in the speedup ratio. 
Within a specific model, an increase in the number of participating devices raises the communication-to-computation ratio, thus magnifying the benefits of our communication optimization.
When compared to SP, Galaxy achieves up to $1.11\times$ performance enhancement. SP requires less synchronous communication than both Galaxy and M-LM, resulting in a smaller speedup ratio. 
However, as SP applies partitioning along the sequence dimension, it necessitates that each device retains a full set of model weights. This requirement is particularly memory-intensive and thus unfriendly to resource-constrained edge devices, as evidenced by frequent OOM issues.

\begin{figure}[t!]
    \setlength{\abovecaptionskip}{-0.1cm}
    \centering
    \includegraphics[width=1\linewidth]{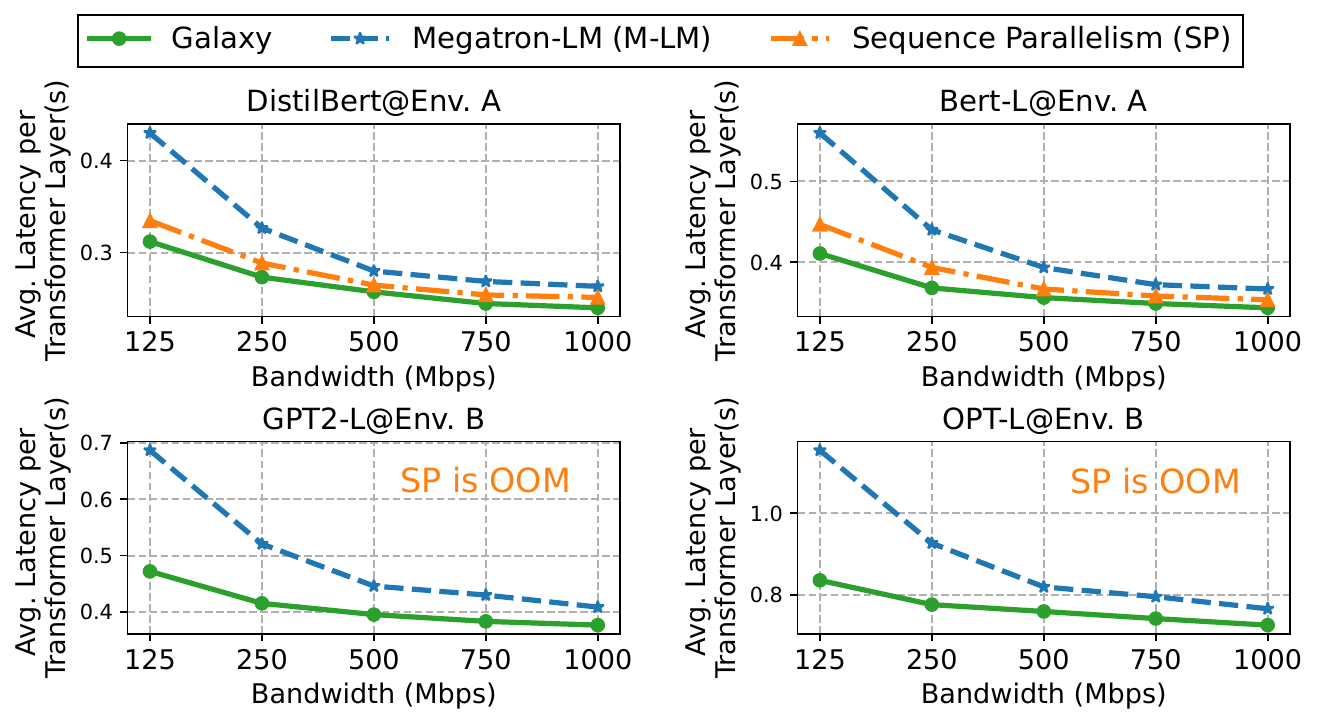}
    % \caption{Galaxy System Overview.}
    \caption{General performance of Galaxy with various network bandwidth.}
    \label{fig:band-lat}
    \vspace{-15pt}
\end{figure}

We further compare Galaxy's performance with baselines under varied network conditions. Using the switcher's traffic control, we simulate five D2D bandwidths to mimic various network conditions at edge. Evaluation results are shown in Fig. \ref{fig:band-lat}.
We observe that in varying network bandwidth conditions, Galaxy consistently exhibits superior performance over baselines, achieving an inference latency reduction of $1.04\times$-$1.45\times$ across diverse models and edge environments.

% Please add the following required packages to your document preamble:
% \usepackage{multirow}
\begin{table}[t!]
\caption{Model Specifications and general performance of Galaxy}
\label{tab:model}
\centering
\begin{tabular}{cccccll}
\hline
\multirow{2}{*}{Model} & \multirow{2}{*}{Layers} & \multirow{2}{*}{Heads} & \multirow{2}{*}{\begin{tabular}[c]{@{}c@{}}Hidden\\ Layer\end{tabular}} & \multirow{2}{*}{\begin{tabular}[c]{@{}c@{}}Edge\\ Env.\end{tabular}} & \multicolumn{2}{l}{Speedup Over} \\ \cline{6-7} 
 &  &  &  &  & M-LM & SP \\ \hline \hline
DistilBert \cite{sanh2019distilbert} & 6 & 12 & 768 & A & $1.37\times$ & $1.08\times$ \\ \hline
\multirow{2}{*}{Bert-L \cite{devlin2018bert}} & \multirow{2}{*}{24} & \multirow{2}{*}{16} & \multirow{2}{*}{1024} & A & $1.36\times$ & $1.09\times$ \\
 &  &  &  & B & $1.38\times$ & $1.11\times$ \\ \hline
\multirow{2}{*}{GPT2-L \cite{radford2019language-gpt2}} & \multirow{2}{*}{36} & \multirow{2}{*}{20} & \multirow{2}{*}{1280} & A & $1.31\times$ & OOM \\
 &  &  &  & B & $1.46\times$ & OOM \\ \hline
\multirow{3}{*}{OPT-L \cite{zhang2022opt}} & \multirow{3}{*}{24} & \multirow{3}{*}{16} & \multirow{3}{*}{2048} & A & $1.26\times$ & OOM \\
 &  &  &  & B & $1.40\times$ & OOM \\
 &  &  &  & C & $1.43\times$ & OOM \\ \hline
\multirow{3}{*}{OPT-XL \cite{zhang2022opt}} & \multirow{3}{*}{32} & \multirow{3}{*}{32} & \multirow{3}{*}{2560} & A & OOM & OOM \\
 &  &  &  & B & OOM & OOM \\
 &  &  &  & C & $1.28\times$ & OOM \\ \hline
\end{tabular}
\vspace{-10pt}
\end{table}

\begin{figure*}[h!]
    \setlength{\abovecaptionskip}{-0.1cm}
    \centering
    \includegraphics[width=1\linewidth]{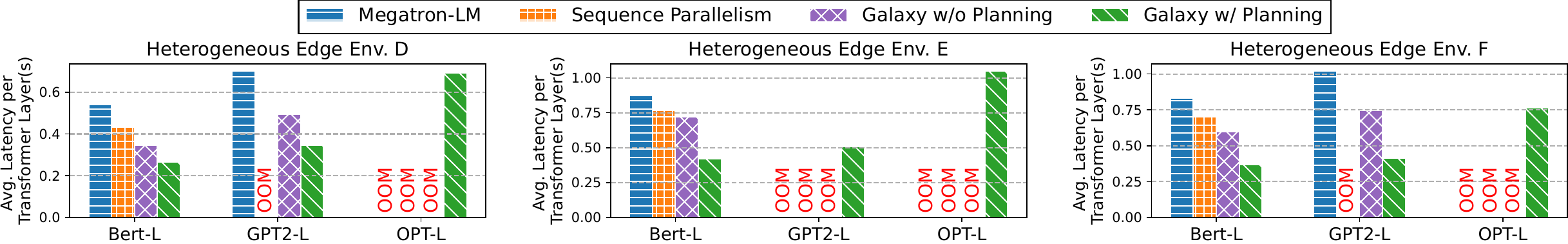}
    % \caption{Galaxy System Overview.}
    \caption{Performance on edge environments with heterogeneous edge devices.}
    \label{fig:hetero}
    \vspace{-15pt}
\end{figure*}

\subsection{Evaluate with Heterogeneous Edge Environments}
We conducted comparisons between Galaxy and baselines within various edge environments (125Mbps), each comprising devices with different computing capacities and memory budgets. The results are demonstrated in Fig. \ref{fig:hetero}.
We observe that Galaxy consistently and remarkably outperforms other state-of-the-art parallelism methods in various heterogeneous edge environments, yielding a substantial inference latency reduction in the range of $1.3\times$ to $2.5\times$. 
Galaxy's superior performance in heterogeneous edge environments derives from its consideration of device heterogeneity, a factor overlooked by M-LM and SP, both tailored for datacenters equipped with homogeneous accelerators. 
% Such oversight results in imbalanced workloads across Transformer inference, leading to resource underutilization.
In addition to device heterogeneity, Galaxy workload planning comprehensively considers the memory budget of edge devices, enabling them to collaboratively accommodate the target model. In contrast, M-LM and SP overlook the memory constraints during parallelism planning, resulting in OOM errors.

\begin{figure}[t!]
    \setlength{\abovecaptionskip}{-0.1cm}
    \centering
    \includegraphics[width=1\linewidth]{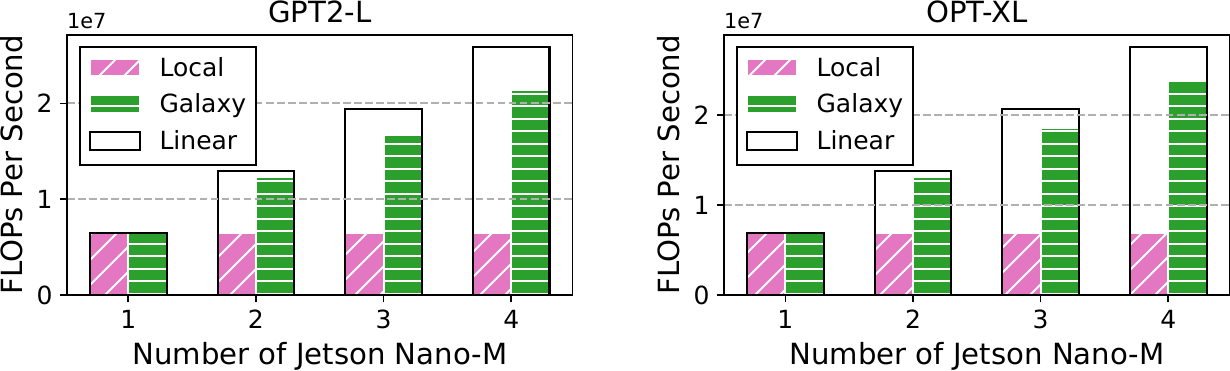}
    \caption{Performance under weak scaling setup.}
    \label{fig:weak-scalability}
    \vspace{-15pt}
\end{figure}

\begin{figure}[t!]
    \setlength{\abovecaptionskip}{-0.1cm}
    \setlength{\belowcaptionskip}{-0.4cm}
    \centering
    \includegraphics[width=1\linewidth]{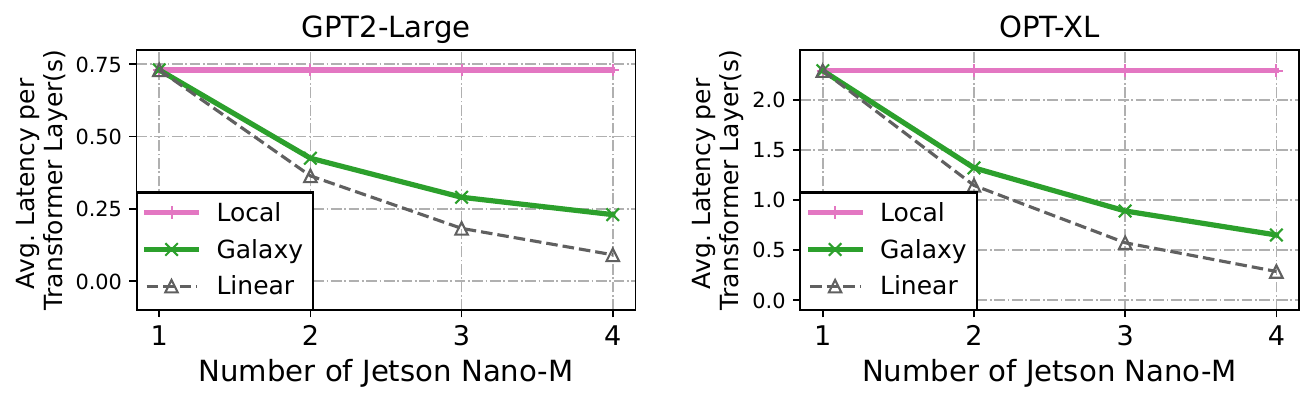}
    \caption{Performance under strong scaling setup.}
    \label{fig:strong-scalability}
    \vspace{-15pt}
\end{figure}

\subsection{Scalability Analysis}
To explore the scalability of Galaxy, we set up both weak  and strong scaling experiments in edge environment C (1000Mbps).
To obviate the impact of OOM errors on our experimental observations, we load and repeatedly perform inference on one single layer, rather than loading entire model.

\subsubsection{Weak Scaling}
In a weak scaling setup, the global workload increases proportionally with the number of devices.
We set a weak scaling with a fixed sequence length of 96 per device (e.g. sequence length is equal to 384 for 4 Jetson Nano-M). 
The overall system's floating-point operations per second (FLOPS) are then evaluated.
% in relation to different number of edge devices participating in the collaborative inference.
As depicted in Fig.\ref{fig:weak-scalability}, we observe excellent scaling performance in both GPT2-L and OPT-XL. Specifically, the GPT2-L case with 4-way (four Jetson Nano-M) HMP can achieve $81\%$ of linear scaling while the OPT-XL case with 4-way can achieve $86\%$ of linear scaling. 

\subsubsection{Strong Scaling}
In a strong scaling setup, the global workload is independent of the number of participating devices. We fix the sequence length to a constant value of 384.
As depicted in Fig. \ref{fig:strong-scalability}, we measure the average inference latency per Transformer layer for a varying number of edge devices.
Galaxy also demonstrates superior scalability under a strong scaling setup. Specifically, Galaxy achieves $3.05\times$ inference latency reduction compared to Local Inference in GPT2-L case, while achieving $3.24\times$ inference latency reduction compare to Local Inference in OPT-XL case.

\subsection{GPU Support}
\label{sec:gpu}
We further evaluate Galaxy's performance in mobile GPUs environments and compare it against baselines. The GPU environment is set up using two Jetson Nanos' onboard GPUs, operating at a locked frequency of 460MHz. The experiments encompass all five Transformer-based models with edge environment A (500Mbps), as shown in Table \ref{tab:GPU-support}.
We observe Galaxy outperforming baselines, achieving an inference latency reduction of $1.12\times$-$1.67\times$ under the GPU environment. Despite the potential underutilization of GPUs for small models like DistilBERT due to Galaxy's communication optimization with matrix tiling, Galaxy still achieves accelerations up to $1.36\times$ compared to baselines.

\begin{table}[t!]
\caption{Inference latency speedup with Mobile GPUs.}
\label{tab:GPU-support}
\centering

\begin{tabular}{c|c|c|c|c|c}
\hline 
\begin{tabular}[c]{@{}c@{}}Speedup\\Over\end{tabular} & DistilBert & Bert-L & GPT2-L & OPT-L & OPT-XL \\ \hline \hline
M-LM       &         $1.36\times$ &    $1.57\times$  &   $1.67\times$   &    $1.58\times$      &   $1.47\times$   \\ \hline
SP         &         $1.12\times$  &   $1.24\times$   &  $1.35\times$  &     $1.26\times$     &   $1.19\times$ \\ \hline
\end{tabular}
\vspace{-15pt}
\end{table}

\section{Related Work}
\noindent\textbf{Collaborative Execution of Transformer.}
Data Parallelism \cite{li2014communication, rajbhandari2020zero} is the most extensively used distributed training approach in datacenters. Pipeline Parallelism is further proposed to conquer the memory issues of training large-scale transformer-based models \cite{narayanan2021efficient,huang2019gpipe}, but suffers from pipeline bubbles. Model Parallelism  simultaneously tackles both memory and bubble issues, and is widely used in both training \cite{narayanan2021efficient, li2021sequence,korthikanti2023reducing} and inference \cite{aminabadi2022deepspeed, yu2022orca, li2023alpaserve} tasks at datacenters. However, few of these approaches are designed for in-situ deep learning at the edge.

\noindent\textbf{In-situ DNN Inference.} 
Pipe-It and Asymo \cite{wang2021asymo,wang2019high} scheduling workload according to the computing power of asymmetry mobile CPU cores to achieve higher throughput.
BlastNet, CoDL and $\mu$layer \cite{ling2022blastnet,jia2022codl,kim2019mulayer} perform a collaborative DNN inference on mobile CPU and GPU concurrently. Band \cite{jeong2022band} coordinates multi-DNN inference on heterogeneous mobile processors. CoEdge, DeepThings, and DeCNN \cite{zeng2020coedge, zhao2018deepthings, du2020model} distribute CNN inference workload over multiple resource-constrained edge devices. However, few of these approaches are designed for Transformer-based models.

% \noindent\textbf{Communication Optimization.}
\noindent\textbf{Communication Optimization for Distributed Deep Learning.}
ZeRO++ \cite{wang2023zero++} utilizes quantized communication to reduce the overhead of communication. Hermes \cite{li2021hermes} applies model structured pruning to achieve communication volume reduction. CoCoNet and ASE \cite{jangda2022breaking,rashidi2021enabling} employ the concept of compute-communication overlap to mitigate communication latency. 
However, few of these approaches are specifically dedicated to model parallelism of Transformer-based models.

\section{Conclusion}
This paper introduces Galaxy, an innovative collaborative in-situ Transformer inference system featuring a hybrid model parallelism architecture, a heterogeneity and memory-budget aware planning algorithm, and a tile-based communication optimization. Our extensive evaluation demonstrates that Galaxy achieves up to $2.5\times$ performance enhancement compare to state-of-the-art approaches.

\bibliographystyle{IEEEtran}
\normalem
\bibliography{reference}

\end{document}